\documentclass[twocolumn,twoside,slac]{revtex4}
\usepackage{graphicx}
\usepackage{fancyhdr}
\begin{document}

\title{Computing System for the Belle Experiment}

%

\author{I.~Adachi, R.~Itoh, N.~Katayama, T.~Tsukamoto, T.~Hibino, M.~Yokoyama }
\affiliation{IPNS, KEK, Tsukuba, Japan}

\author{L.~Hinz, F.~Ronga}
\affiliation{IPHE, Universit\'e de Lausanne, Lausanne, Switzerland}

\begin{abstract}
We describe the offline computing system of the Belle experiment, consisting
of a computing farm with one thousand IA-32 CPUs. 
Up to now, the Belle experiment has accumulated more than 120 fb$^{-1}$ of data, 
which is the world largest $B\bar{B}$ sample at the $\Upsilon(4S)$ energy. 
The data have to be processed with a single version of reconstruction software 
and calibration constants to perform precise measurements of $B$ meson decays. 
In addition, Monte Carlo samples three times larger than the real beam data 
are generated. To fullfill our computing needs, 
we have constructed the computing system 
with 90(300) quad(dual) CPU PC servers from multiple vendors as a central
processing system. 
The details of this computing system and performance of data
processing with the current model are presented.

\end{abstract}

\maketitle

\thispagestyle{fancy}


\section{Belle Experiment}
\label{sect:BELLE_exp}

The Belle experiment\cite{belle-exp} is the KEK $B$-factory project with 
an asymmetric e$^+$e$^-$ collision to explore CP violation in $B$ meson system.
The data-taking stared from June 1999, and
the accelerator( KEKB )\cite{kekb} gradually improved its performance. The integrated
luminosity logged by the Belle detector has reached 120 fb$^{-1}$ in March 2003.
This corresponds to the fact that more than 120 M $B\bar{B}$ pairs have been
recorded in our tape storage. This is obviously the largest $B\bar{B}$ data sample
at the $\Upsilon(4S)$ energy region in the world.

The KEKB accelerator is still updating their luminosity records 
thanks to excellent operations.
The integrated luminosity per day has been approaching 500 pb$^{-1}$.

From the computing model's point of view, a large data sample is a challenging issue for
CPU power as well as for storage.

In daily event processing, we have to process beam data without any delay 
for online data acquisition. To do so, we need enough CPU power and
a stable DST production.

Furthermore, we have to reprocess the entire data sample whenever
we have a major update of reconstruction codes as well as calibration constants. 
Here, the whole event data have to be reconstructed from rawdata 
using the same version of the codes and constants
to controll the systematic errors in user analyses.

Considering the large amount of accumulated beam data, it is unrealistic 
for users to make their analyses in an efficient way using the whole data 
sets. Thus, the BELLE data processing also incorporates the data skimming, 
leading to reduced data files based on a first event selection and called 
physics skims.

For the Monte Carlo( MC ) data, we require large amount of statistics, at least
3 times larger than beam data to evaluate systematic effect 
related to the detector acceptance, event reconstruction and so on.

In this paper, we first introduce 
the BELLE computing software(section \ref{sect:software_tools}) 
and system(section \ref{sect:BELLE_computing_system}) including PC farms 
and its upgrade(section \ref{sect:pc_farms}). 
Then, in section \ref{sect:DST_production}, we will explain 
the scheme of the DST production/reprocessing and 
mention how MC events have been produced(section \ref{sect:MC_production}).
Finally summary and future plan will be given.

\section{Software Tools}
\label{sect:software_tools}

We describe here the core software of event processing 
which is based on ``home-made'' tools.

\subsection{B.A.S.F. Framework}

B.A.S.F.( Belle AnalySis Framework ) is a unique framework for us from DAQ to final
user analyses. 
It is composed by a set of modules written in C++, 
each one having at least the following structure: a beginning of run, 
an event processing and an end of run function. 
Histograming is also included in the structure of modules.
These modules are compiled as shared objects 
and are dynamically plugged in when B.A.S.F. runs 
as shown in Figure~\ref{fig:event_flow}.
This framework utilizes event-by-event parallel processing on SMP using a fork function.

\subsection{Data Management}

The data transfer between modules and I/O is managed by PANTHER. This is based upon 
a bank system composed of tables. The contents of tables
are defined  in ASCII format before loading modules
and user has to include this file in the code as header files.
In each tables, one can pick up any value corresponding to each attribute and
pointer which relates one table to the other. This contains compression capability
using zlib utility.
PANTHER is only a data management system in the Belle experiment. From rawdata
to user analyses, any stage of events can be consistently handled only with PANTHER.

The typical event size is 35 KB for rawdata and for 60 KB for reconstructed DST data.
The DST data contains a lot of information of results from each detector analyses.
The quantity of information stored in these DST files is usually too large for a 
standard physics analysis.
To reduce the data size, 
we produce compact format of the DST data(``mini-DST''), where the size is 
12 KB per hadronic event.

\subsection{Reconstruction Library and Simulation}

The detector calibration constants, used to process data, 
are stored in a PostgreSQL\cite{postgres} database.
At KEK, we have two database servers and one is mirrored to the other.
In each institution, there is a database server of which the substance is periodically 
imported from the KEK main server.

\begin{figure}
\includegraphics[width=65mm]{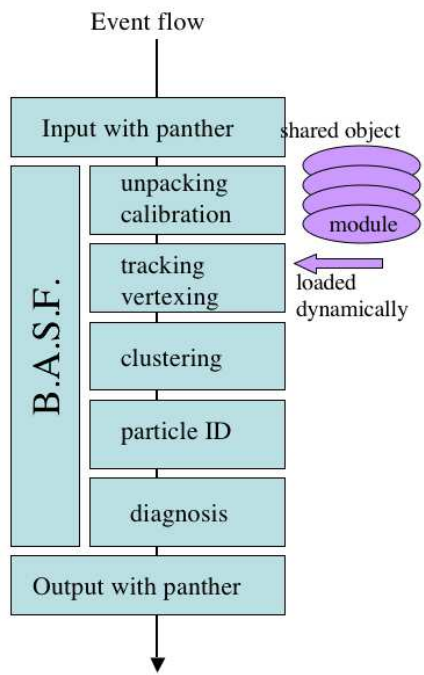}
\caption{Schematic view of event flow.}
\label{fig:event_flow}
\end{figure}

The event flow of DST production is schematically shown in Figure~\ref{fig:event_flow}.
Each step like calibration and unpacking is performed by a set of modules which are
loaded on the B.A.S.F. platform. 
The reconstructed data are transfered in the PANTHER format
between different modules. At the end, the data are compressed.

The update of the library is usually done a couple of times in a year. 
In the last year, we had a major update in April, where the tracking software 
for low momentum region was improved.
Once a new library is released, all events taken 
so far are again processed( reprocessing ) from rawdata in order to produce 
new DST and mini-DST file.

The simulation is performed with GEANT3 packages\cite{geant} 
by adding an interface to B.A.S.F.
Then, MC data follow the same treatment as the real data, including the 
detector calibration constants database. For every significant library changes, 
as in spring 2002, the MC data samples are generated using the same version of 
BASF library than real data.

\section{Belle Computing System}
\label{sect:BELLE_computing_system}

\begin{figure*}[t]
\centering
\includegraphics[width=135mm]{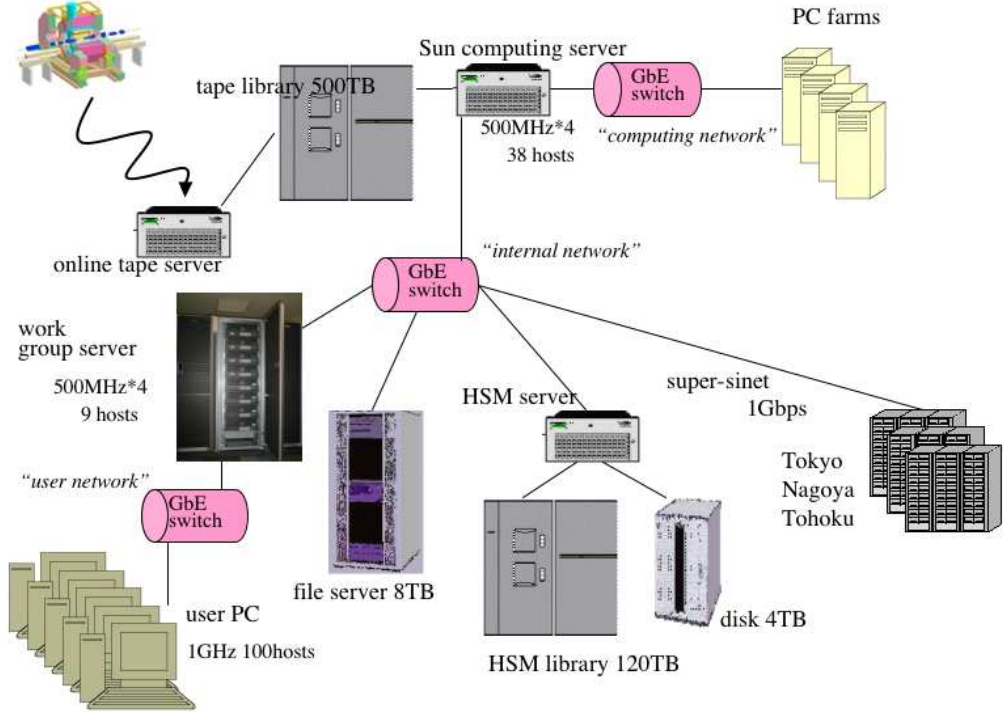}
\caption{Belle computing system} 
\label{fig:belle_computing}
\end{figure*}

Figure~\ref{fig:belle_computing} is an overview of our computing system which consists of the three 
principal networks. The first one is connected between computing servers for batch jobs and
the PC farms( described in the next section ) for
DST/MC productions. This system is composed of 38 Sun hosts as computing machines
linked to PC farms via a gigabit ethernet switch( ``computing network'' ).   
These Sun hosts are operated under the LSF batch qeueing utility. 
20 Sun hosts out of 38 are equipped with 2 DTF2 tape drives each and are 
connected to the SONY tape robotic library of 500 TB capacity.
The PC farms are used for the DST/MC productions.
Rawdata from the Belle detector are sent to an online tape server
linked to the tape drives, by which rawdata are written onto a DTF2 tape.

The second network is used for user analyses and data storage system. The 9 work group
servers are connected to a swiching device of a gigabit ehternet called ``internal network''.
The hierarchy mass storage( 120 TB capacity ) with the 4 TB staging disk in addition to
the 8 TB file servers are also connected. The other network switch( ``user network'' ) links
between the work group servers and 100 user PC's. Everybody can use each PC for her/his analysis,
or login to one of the work group servers and analyze data interactively. For the batch system,
user can submit batch jobs to the computing servers from the work group servers.
The CPU's are summarised in Table~\ref{tb:Belle-CPU}. 

\begin{table}
\begin{center}
\begin{tabular}{|c|c|c|c|c|c|} \hline
host              & processor &  clock    & \#CPU's & \#nodes    \\ \hline  
computing server  & sparc     & 500MHz    &    4    &      38    \\ \hline
work group server & sparc     & 500MHz    &    4    &       9    \\ \hline
user PC           & P3        & 1GHz      &    1    &     100    \\ \hline
\end{tabular}
\caption{\small Summary of CPU's for the Belle computing system. }
\label{tb:Belle-CPU}
\end{center}
\end{table}

\section{PC Farms}
\label{sect:pc_farms}

\begin{figure}
\includegraphics[width=65mm]{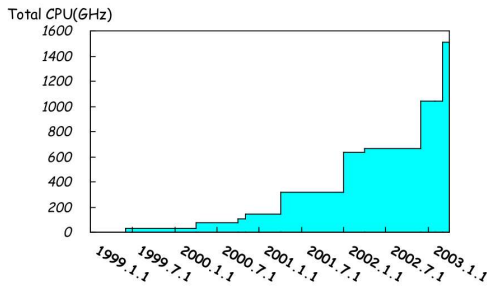}
\caption{History of the CPU upgrade since 1999.}
\label{fig:PCstatus}
\end{figure}

Two requirements on our computing system
in order to achieve physics goal are addressed:

\begin{itemize}

\item
Whole beam data have to be reprocessed with a single
version of the reconstruction code and constants to provide a consistent dataset.
Our aim is to do it within three months, which enables us to obtain physics results promptly.

\item
Quantity of MC sample to be used for data analyses must be at least 3 times larger than
that of beam data size. 
We need this requirement, for instance, to study the systematic errors.

\end{itemize}

To satisfy these requirements, we have added more PC farms 
in our computing model described in the previous section and
boosted up CPU power for the DST/MC production.
Figure~\ref{fig:PCstatus} shows total CPU of the PC farms 
installed in the system as a function of time, where 
a vertical axis represents CPU power in unit of GHz, which
is estimated as a product of processor speed, \# of processors and \# of PC nodes.
We have started purchasing PC farms since June 1999. Then, PC farms
have been gradually installed as we accumulated more integrated luminosity.

Table~\ref{tb:PC-CPU} summarises our PC farms. As we can seen,
our system is heterogeneous from various vendors. 
In total, CPU power of 1508GHz is equipped in our system.

\begin{figure*}[t]
\centering
\includegraphics[width=135mm]{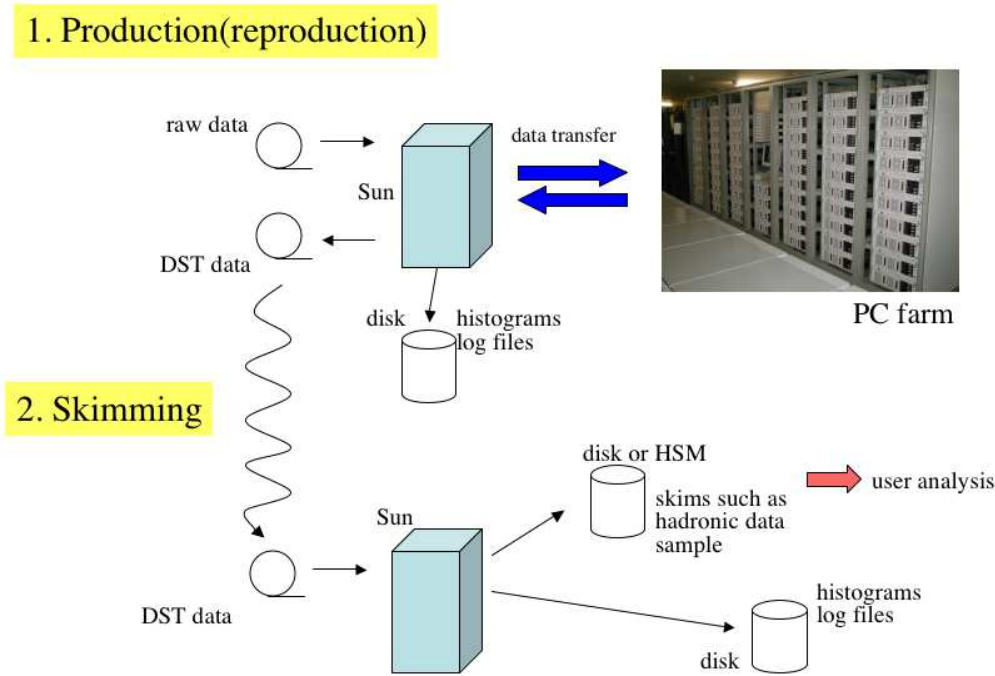}
\caption{A scheme of the DST production(reproduction).}
\label{fig:prod_scheme}
\end{figure*}

\begin{table}
\begin{center}
\begin{tabular}{|c|c|c|c|c|c|} \hline
vendor & processor &  clock    & \#CPU's & \#nodes  & total CPU    \\ \hline  
Dell   & P3        & 500MHz    &    4    &      16  &     32GHz    \\ \hline
Dell   & P3        & 550MHz    &    4    &      20  &     44GHz    \\ \hline
Compaq & P3        & 800MHz    &    2    &      20  &     32GHz    \\ \hline
Compaq & P3        & 933MHz    &    2    &      20  &     37GHz    \\ \hline
Compaq & Intel Xeon& 700MHz    &    4    &      60  &    168GHz    \\ \hline
Fujitsu& P3        & 1.26GHz   &    2    &     127  &    320GHz    \\ \hline
Compaq & P3        & 700MHz    &    1    &      40  &     28GHz    \\ \hline
Appro  & Athlon    & 1.67GHz   &    2    &     113  &    377GHz    \\ \hline
NEC    & P3        & 2.8GHz    &    2    &      84  &    470GHz    \\ \hline\hline
Total  &           &           &         &     500  &   1508GHz    \\ \hline 
\end{tabular}
\caption{\small A breakdown of the PC farm CPU's. }
\label{tb:PC-CPU}
\end{center}
\end{table}

In all PC farms, RedHat linux systems of version 6 or 7\cite{redhat} have been installed
and all of Belle utilities are implemented.

\section{DST Production and Reprocessing}
\label{sect:DST_production}

\begin{figure}
\includegraphics[width=65mm]{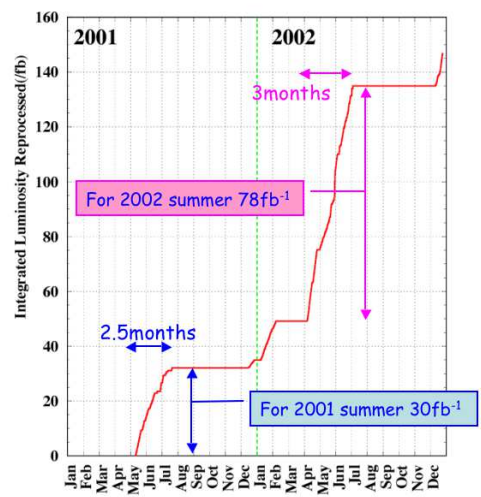}
\caption{A history of reprocessing done in 2001 and 2002.}
\label{fig:reprocess_status}
\end{figure}

A schematic drawing of the production scheme is shown in Figure~\ref{fig:prod_scheme}. 
The first step( ``production'' ) is performed in the following way.
One of the computing servers of Sun hosts is employed as a tape server, where two
DTF2 tapes, one for rawdata and the other for DST data, are mounted.
Rawdata stored in a DTF2 tape is read on the tape server and these data are distributed
over all hosts in the PC farm, in which the event processing is performed. 
After the event reconstruction,
the processed data are sent back to the tape server 
and they are recorded in the output tape as DST.

In the next step(``skimming''), the DST tape is read at the different tape server and, based upon
event classification criteria they are categorized into several physics skims. Then,
only events of our interest are written onto disks. These output skims are
starting point of individual analyses.

For the reprocessing, one remarkable feature in the skimming stage is that it is possible 
to write the output data onto a disk located outside KEK. For instance, we can use a disk 
at Nagoya, 350 km away from KEK.
This disk is mounted to the file server using NFS 
via the gigabit network of SuperSINET\cite{supersinet}.

The performace of the reprocessing we have done since 2001 is 
shown in Figure~\ref{fig:reprocess_status}. In 2001, the first turn-around of the reprocessing 
has started in April after all of the reconstruction programmes together with calibration constants
had been fixed. The whole data of 30 fb$^{-1}$ has been reprocessed before August and were
used for the analyses for the 2001 summer conferences. The next turn-around was made last year.
The library for the reconstruction has been frozen in the beginning of April 2002, and the reprocessing
has started again. Here, all of the data taken from 1999 to July 2002 have been reconstructed
to produce a unique and a consistent data set for physics analyses.
As can be seen in the Figure~\ref{fig:reprocess_status}, 
our reprocessing was successfully done within a couple of months.
For reprocessing performed in 2002, compared to 2001 beam data, 
the data size is increased by a factor of 2.5, however
the upgraded CPU power by adding PC farms allows us to reconstruct all the 
rawdata of 78 fb$^{-1}$ in 3 months.

\begin{table}
\begin{center}
\begin{tabular}{|c|c|c|c|c|c|} \hline
module crash   & $< 0.01$ \%   \\ \hline
tape I/O error & 1 \%          \\ \hline
process comunication error & 3 \% \\ \hline
network trouble/system error & negligible \\ \hline
\end{tabular}
\caption{\small A summary of the failure rate in reprocessing. }
\label{tb:fail_rate}
\end{center}
\end{table}

The failure rate of the reprocessing is tabulated in Table~\ref{tb:fail_rate}.
The main trouble comes from the comminication error among PC hosts, where  
possible improvement of the signal handling of each PC could be expected to reduce
this error. However, our total error rate is still small enough 
and the whole comuputing system has been efficiently working.

\section{MC Production}
\label{sect:MC_production}

\begin{figure}
\includegraphics[width=65mm]{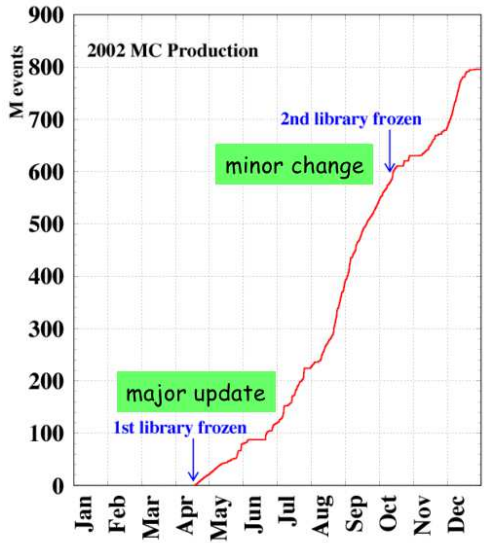}
\caption{MC events produced in 2002.}
\label{fig:MC_status}
\end{figure}

The MC data consists of 3 type of physics events of $B^0\bar{B^0}$, $B^+B^-$
and continuum. 

We produce these MC data using the real beam conditions, 
like beam pipe background and interaction point(IP) position. 
These conditions depending on the run number, we called this procedure a ``run-by-run''  
production of MC data. 
Moreover, as it was already mentionned, we generate MC data with 3 times larger 
statistics than real data.

Figure~\ref{fig:MC_status} represents how we have produced MC events
in 2002. Major update of the simulation and reconstruction software
was done in April, and since then we have started
the MC production. In October, a minor change of the simulation code in order to match
to the modification of the beam triggering scheme was made.

The PC farms were shared between the DST and the MC production and the allocation 
of the CPU power can be easily modified according to the situation.

\begin{figure}
\includegraphics[width=65mm]{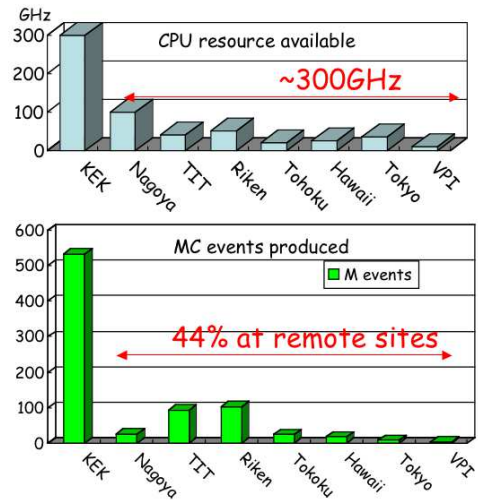}
\caption{CPU used for the MC production(top) and \# of events produced(bottom) in each site.}
\label{fig:MC_remote}
\end{figure}

For the MC production, the computing resources at remote sites have been actively used. The upper
Figure~\ref{fig:MC_remote} indicates that total CPU power at remote sites is about 300 GHz.
This is comparable to that of KEK availble for the MC production. 
The MC events produced outside KEK amounts to 44 \% of all of events produced(shown in
Figure~\ref{fig:MC_remote}).
These MC data are basically sent to KEK via network and are saved in the disk.
In case that the disk does not have enough space, MC events were copied onto DTF2 tape instead. 
The MC data tapes are released in the tape library and
then user can access these data in batch jobs.
The typical size corresponds to around 6 TB in 6-months MC production.
More remote institutions are expected to join in producing the MC events this year.

\section{Summary and Future Pan}
\label{sect:conclusion}

The Belle computing system has been operated in an efficient way so that
we have successfully reprocess more than 250 fb$^{-1}$ of real beam data so far. 
For the MC data, event samples 3 times larger than beam data have been produced 
at KEK and at remote sites. 

A higher luminosity $B$-factory machine at KEK( superKEKB ) is being proposed as an upgrade plan of 
the present experiment\cite{superKEKB}. This plan aims to achieving 
$10^{35}$ cm$^{-2}$sec$^{-1}$ or more luminosity, 
corresponding to a data size of 1 PB for 1-year operation. 
To handle this amount of data, 
we may have to introduce Grid technology, for instance, for efficient usage of resources
at remote sites. The new computing model for the superKEKB experiment will be
presented in letter of intent submitted in the end of 2003.

\begin{acknowledgments}
The authors wish to thank members at KEK Computing Reseach Centre 
for their support.
\end{acknowledgments}


\end{document}